\documentclass[traditabstract]{aa}

\usepackage{amssymb}
\usepackage{amsmath}
\usepackage{txfonts}
\usepackage{graphicx}
\usepackage{colortbl}
\usepackage[usenames]{xcolor}
\usepackage[bookmarks, colorlinks, breaklinks]{hyperref}
\hypersetup{linkcolor=blue,citecolor=blue,filecolor=black,urlcolor=blue} 
\usepackage{natbib}
\bibpunct{(}{)}{;}{a}{}{,} % to follow the A&A style
\defcitealias{2000A&A...353..117M}{M00}
\defcitealias{2002A&A...384..866M}{M02}

\begin{document}

\title{Large and small-scale structures and the dust energy balance problem in spiral galaxies}

\titlerunning{Dust energy balance problem}

\author{
W. Saftly\inst{\ref{UGent}} \and
M. Baes\inst{\ref{UGent}} \and
G. De Geyter\inst{\ref{UGent}} \and
P. Camps\inst{\ref{UGent}} \and
F. Renaud\inst{\ref{USurrey}} \and
J. Guedes\inst{\ref{TAG}} \and
I. De Looze\inst{\ref{UGent},\ref{UK}}
}

\authorrunning{W. Saftly et al.}

\institute{ Sterrenkundig Observatorium, Universiteit Gent, Krijgslaan
  281, B-9000 Gent, Belgi\"e\label{UGent} 
  \and
  Department of Physics, 14 BC 03, University of Surrey, Guildford GU2 7XH, United Kingdom\label{USurrey}
  \and
   Teralytics AG, Zollstrasse 62, 8005 Zurich, Switzerland\label{TAG}
   \and
   Institute of Astronomy, University of Cambridge, Madingley Road, Cambridge, CB3 0HA, United Kingdom\label{UK}
}

\date{\today}
\abstract{The interstellar dust content in galaxies can be traced in extinction at optical wavelengths, or in emission in the far-infrared. Several studies have found that radiative transfer models that successfully explain the optical extinction in edge-on spiral galaxies generally underestimate the observed FIR/submm fluxes by a factor of about three. In order to investigate this so-called dust energy balance problem, we use two Milky Way-like galaxies produced by high-resolution hydrodynamical simulations. We create mock optical edge-on views of these simulated galaxies (using the radiative transfer code SKIRT), and we then fit the parameters of a basic spiral galaxy model to these images (using the fitting code FitSKIRT). The basic model includes smooth axisymmetric distributions along a S\'ersic bulge and exponential disc for the stars, and a second exponential disc for the dust. We find that the dust mass recovered by the fitted models is about three times smaller than the known dust mass of the hydrodynamical input models. This factor is in agreement with previous energy balance studies of real edge-on spiral galaxies. On the other hand, fitting the same basic model to less complex input models (e.g. a smooth exponential disc with a spiral perturbation or with random clumps), does recover the dust mass of the input model almost perfectly. Thus it seems that the complex asymmetries and the inhomogeneous structure of real and hydrodynamically simulated galaxies are a lot more efficient at hiding dust than the rather contrived geometries in typical quasi-analytical models. This effect may help explain the discrepancy between the dust emission predicted by radiative transfer models and the observed emission in energy balance studies for edge-on spiral galaxies. }
\keywords{ radiative transfer \and Galaxies: ISM \and dust,extinction \and hydrodynamics \and methods: numerical}
\maketitle

\section{Introduction}
\label {Introduction.sec}

Interstellar dust grains in galaxies play a special role in producing and processing radiation. They efficiently absorb and scatter ultraviolet (UV), optical and near-infrared (NIR) photons and then reradiate the absorbed energy in the infrared and submm wavelength range. As a result, the presence of dust influences the apparent view of a galaxy and apparent structural parameters such as scale lengths, surface brightnesses, luminosities and colours. The effects of dust attenuation on these parameters are nontrivial and sometimes counterintuitive \citep{1994ApJ...432..114B, 2004A&A...419..821T, 2004ApJ...617.1022P, 2010MNRAS.403.2053G, 2013A&A...553A..80P, 2013A&A...557A.137P}. A major source of complexity is the relative geometry between stars and dust in a galaxy: the same amount of dust can give rise to completely different levels of attenuation depending on the star-dust geometry \citep{1992ApJ...393..611W, 2001MNRAS.326..733B}. Before we can correct for the attenuating effects in a galaxy, we need to understand the total amount and spatial distribution of interstellar dust.

Broadly speaking, the dust in galaxies can be traced in two ways. The first way is by studying the thermal emission of the dust at far-infrared (FIR) and submm wavelengths. We can determine the total dust mass of a galaxy with a reasonable accuracy by fitting a modified blackbody or more advanced models to the FIR/submm observed spectral energy distribution. Especially after the launch of the {\it Herschel} Space Observatory \citep{2010A&A...518L...1P}, a wealth of FIR/submm data has become available that allowed different teams to determine dust masses in thousands of galaxies \citep[e.g.,][]{2011MNRAS.417.1510D, 2012ApJ...745...95D, 2013MNRAS.436.2435S, 2014A&A...565A.128C}. Due to the diffraction limit in the FIR/submm and the limited diameter of space observatories, detailed investigations of the distribution of the dust remains difficult, except for the most nearby galaxies \citep[e.g.,][]{2010A&A...518L..67K, 2012A&A...546A..34F}.

An alternative method, which does not suffer from the poor spatial resolution affecting FIR/submm observations, consists of carefully modelling the attenuation effects of the dust on the stellar emission in the optical window using advanced radiative transfer techniques. Today, a range of powerful 3D radiative transfer codes is available \citep[for an overview, see][]{2013ARA&A..51...63S}, and the fitting of the radiative transfer models to data is being optimised beyond the elementary chi-by-eye \citep{1997A&A...325..135X, 2005A&A...434..167S, 2007A&A...471..765B, 2012ApJ...746...70S, 2013A&A...550A..74D, 2014MNRAS.441..869D}.

Traditionally, detailed radiative modelling has been applied mainly to edge-on spiral galaxies, as the dust is then clearly visible as prominent dust lanes in optical images. Most studies focus on a single galaxy \citep[e.g.,][]{1987ApJ...317..637K, 1997A&A...325..135X, 1998A&A...331..894X, 2000A&A...362..138P, 2011A&A...527A.109P, 2010A&A...518L..39B, 2012MNRAS.427.2797D, 2012MNRAS.419..895D, 2012ApJ...746...70S} or modest sets of edge-on spiral galaxies \citep{1999A&A...344..868X, 2007A&A...471..765B, 2011ApJ...741....6M, 2014MNRAS.441..869D}.
 
A comparison of the extinction and the FIR/submm emission, i.e.\ a study of the dust energy balance, gives the strongest constraints on the dust content of spiral galaxies. Dust energy balance studies of edge-on spiral galaxies reveal a discrepancy between the FIR/submm emission predicted by radiative transfer models and the observed emission. Although the radiative transfer models explain the optical extinction very well, they typically underestimate the observed dust emission by a factor of about three \citep{2000A&A...362..138P,2001A&A...372..775M, 2004A&A...425..109A, 2005A&A...437..447D, 2010A&A...518L..39B, 2012MNRAS.427.2797D, 2012MNRAS.419..895D}. This has become generally known as the dust energy balance problem in edge-on spiral galaxies.

Two broad scenarios have been proposed to explain this problem. A first possibility is that FIR/submm emissivity, standardly taken from semi-empirical dust models based on the interstellar medium (ISM) in the Milky Way, is underestimated by a factor of a few. This idea was advocated by \citet{2004A&A...425..109A} and \citet{2005A&A...437..447D}, and is corroborated by the large range of empirical values for emissivity circulating in the literature. However, based on a detailed study of the edge-on spiral UGC\,4754, \citet{2010A&A...518L..39B} argued that this possibility cannot explain the discrepancy. Indeed, they found an incompatibility between the absorbed stellar luminosity as obtained from the radiative model fits to the optical data and the emitted dust luminosity in the FIR/submm that cannot be lifted by just increasing the value of the dust emissivity. 

The alternative scenario to explain the dust energy balance problem in spiral galaxies is geometrical in nature. As already indicated, the relative star-dust geometry in a galaxy is extremely important for the amount of attenuation. If a sizeable fraction of the dust is distributed in such a way that it hardly attenuates the bulk of the starlight, one could easily underestimate the amount of dust from radiative transfer modelling. At the same time, this dust can still contribute substantially to the observed FIR/submm emission. How exactly this additional dust would have to be distributed is subject to debate, and various options have been proposed. 

One option is a very thin and dense dust disc next to the thicker dust disc that is responsible for the dust lane in the optical \citep{2000A&A...362..138P, 2011A&A...527A.109P, 2001A&A...372..775M, 2007MNRAS.379.1022D}. For example, \citet{2014ApJ...795..136S} find a very thin dust disk in certain edge-on spiral galaxies. On the other hand, \citet{2005A&A...437..447D} showed that such a disc would have an observational signature at NIR wavelengths, which is at odds with the observations of the prototypical edge-on spiral galaxy NGC\,891. 

A second option is that most of the dust is locked up in so-called ``clumps'', a rather vaguely defined term used in the radiative transfer community to indicate any form of inhomogeneities relative to a smooth medium. One can consider large-scale inhomogeneities such as bars and spiral arms, as well as small-scale structures such as dusty molecular clouds with or without embedded young stars. Thanks to the advancement of 3D dust radiative transfer, the effect of a non-homogeneous multi-phase dusty medium has been investigated by various teams \citep[e.g.,][]{1996ApJ...463..681W, 2000ApJ...528..799W, 1998A&A...340..103W, 1999ApJ...523..265V, 2000MNRAS.311..601B, 2003MNRAS.342..453H, 2006ApJ...636..362I}. While the dusty ISM in spiral galaxies is far from smooth and homogeneous, the radiative transfer models being fitted to the optical images in energy balance studies are usually smooth and axisymmetric. This simplification may very well affect the results of these studies.
 
Whether or not an inhomogeneous distribution of the dust, on small and/or on large scales, is a possible solution for the dust energy balance problem is still an open question. Several studies have attempted to address this issue. Concerning the role of spiral arms, the work of \citet[][hereafter \citetalias{2000A&A...353..117M}]{2000A&A...353..117M} is the most advanced study. They set up a suite of galaxy models with an analytical spiral structure, created mock edge-on optical images for these galaxies, and subsequently fitted these simulated observations with a smooth, axisymmetric radiative transfer model. The fitted models were surprisingly accurate; essentially all parameters of the input model (dust mass, inclination, scale parameters of stars and dust, etc.) were properly recovered. In a similar way, \citet[][hereafter \citetalias{2002A&A...384..866M}]{2002A&A...384..866M} investigated the effect of clumps or small-scale inhomogeneities: rather than models with a spiral perturbation they adopted the clumpy disk galaxy models of \citet{2000MNRAS.311..601B} as input models. They concluded that, for the range of clumpy distributions they considered, the neglect of clumping results in a systematic underestimate of the dust mass. The underestimate was found to be never larger than 40\%, however.

These results seem to cast doubts on the ability of inhomogeneities such as spiral arms or large molecular clouds to provide an answer to the dust energy balance problem. However, one needs to take into account that the input models used in the studies by \citetalias{2000A&A...353..117M} and \citetalias{2002A&A...384..866M} were still relatively well-behaved. They were constructed using theoretical perturbations on essentially the same exponential discs as those used in the fitting, and still featured a rather symmetric and regular geometry. They are therefore a relatively poor representation for real galaxies, which are generally characterised by a much larger degree of geometrical complexity and asymmetry. 

The goal of this paper is to investigate whether a complex and inhomogeneous dusty ISM could provide an answer to the dust energy balance problem. In Sect.~\ref{BasicInput.sec} we start with basic input models similar to those used by \citetalias{2000A&A...353..117M} and \citetalias{2002A&A...384..866M}, and we verify that our fitting procedure recovers similar results. In Sect.~\ref{HydroInput.sec} we apply the same approach to galaxy snapshots produced by state-of-the-art hydrodynamic simulations. We consider two models from different simulations, both of which show a complex and asymmetric geometry similar to our own Milky Way. In Sect.~\ref{Discussion.sec} we discuss and interpret the results, and a summary is presented in Sect.~\ref{Conclusion.sec}.

\section{Basic input models}
\label{BasicInput.sec}

Before we set out to model mock images from hydrodynamical simulations and compare the results to previous work, we first need to investigate whether we can reproduce the results of this previous work. In particular, we should verify that our fitting routine can reproduce the results of \citetalias{2000A&A...353..117M} and \citetalias{2002A&A...384..866M} for basic input models. This is not as obvious as it may seem. Indeed, while we follow a similar approach, there are potentially relevant differences in the modelling mechanisms. For example, the radiative transfer code used by \citetalias{2000A&A...353..117M} and \citetalias{2002A&A...384..866M} adopts the so-called method of scattered intensities \citep[for details, see][]{1987ApJ...317..637K, 1997A&A...325..135X, 2001MNRAS.326..722B}, whereas the fundamental algorithm in our SKIRT code is the Monte Carlo method. Perhaps more importantly, while \citetalias{2000A&A...353..117M} and \citetalias{2002A&A...384..866M} fit each model to a single image, we use oligochromatic fitting, i.e.\ we fit a single model simultaneously to images in the five SDSS optical bands. This approach can help to eliminate degeneracies in the parameter space \citep{2014MNRAS.441..869D}. Hence, we set up two test cases that are very similar to the basic input models explored by \citetalias{2000A&A...353..117M} and \citetalias{2002A&A...384..866M}.

\subsection{The input models}

The first step in our modelling is the setup of the models, i.e., the choice of the basic model from which to start and the perturbations that we apply this model. For the underlying model, we assume a smooth and axisymmetric model, consisting of a double-exponential stellar disc, a flattened S\'ersic bulge, and a double-exponential dust disc \citep[see also, e.g.,][]{1999A&A...344..868X, 2007A&A...471..765B, 2011ApJ...741....6M}. For the parameters of the underlying model, we use the average values obtained by fitting 10 real galaxies with FitSKIRT from the CALIFA survey \citep[see Table~4 in][]{2014MNRAS.441..869D}. 

First, we consider a spiral structure perturbation, similar to the models by \citetalias{2000A&A...353..117M}. Both the stars in the disk and the dust are perturbed into a logarithmic spiral arm pattern. For the parameters of the spiral arm perturbation, we select an average model also considered in \citetalias{2000A&A...353..117M}. There are two spiral arms, the weight of the spiral perturbation is 30\% for the stars and 40\% for the dust, and the spiral arm pitch angle is 20~degrees.

Secondly, we set up an idealised model for a clumpy disc galaxy similar to \citetalias{2002A&A...384..866M}. In this case, we divide the dusty medium into a smooth and a clumpy component. We take a different approach from \citet{2000MNRAS.311..601B}, who adopted the two-phase medium algorithm explored by many authors \citep{1996ApJ...463..681W, 2000ApJ...528..799W, 2001ApJ...548..150M, 1998A&A...340..103W, 2012MNRAS.420.2756S}. Instead, we deposit half of the total dust mass in 10,000 individual spherical clumps, each with an outer radius of 300~pc and an internal cubic spline kernel distribution \citep{1981csup.book.....H}. Each of these clumps is positioned at a random location chosen according to the underlying smooth dust density field \citep[see also][]{2014A&A...571A..69D, 2015A&C.....9...20C}. 

\subsection{Creation of mock images}

The second step in our analysis is the creation of mock images for these two models. In particular, we need images in the SDSS {\em{ugriz}} bands and in the edge-on orientation, so that they can be used subsequently for radiative transfer modelling. The mock images were created with SKIRT, a 3D Monte Carlo dust radiative transfer code designed to simulate continuum radiation transfer in dusty astrophysical systems \citep{2003MNRAS.343.1081B,2011ApJS..196...22B, 2015A&C.....9...20C}. It supports multiple anisotropic scattering, absorption and re-emission by interstellar dust (including stochastically heated grains), and can create spectral energy distributions and images at arbitrary wavelengths and from arbitrary points of view. SKIRT is publicly available from a GitHub code repository.\footnote{SKIRT code repository: \url{https://github.com/skirt/skirt}, SKIRT documentation: \url{http://www.skirt.ugent.be}}

\subsection{Radiative transfer modelling}

\begin{figure*}
\centering
\includegraphics[width=\textwidth]{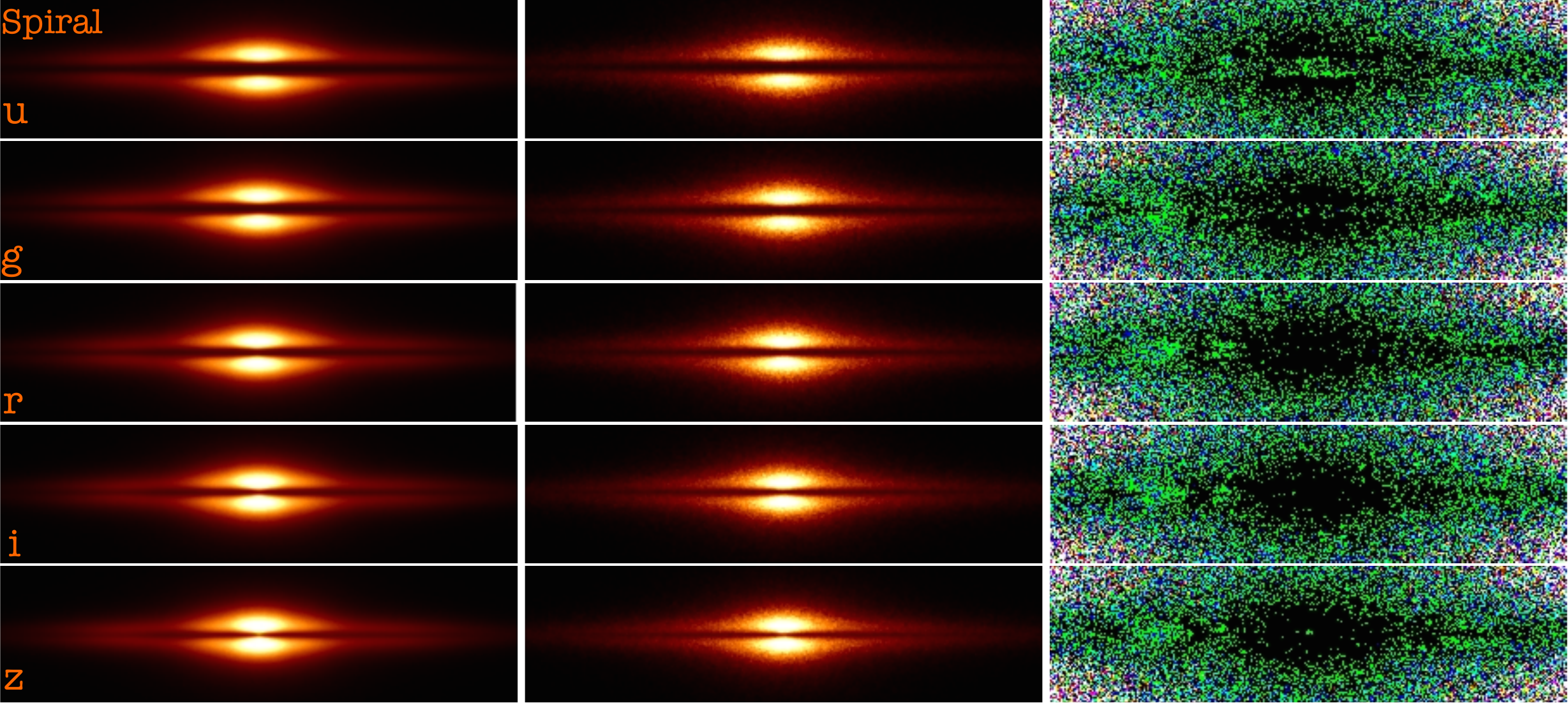}\\[1em]
\includegraphics[width=\textwidth]{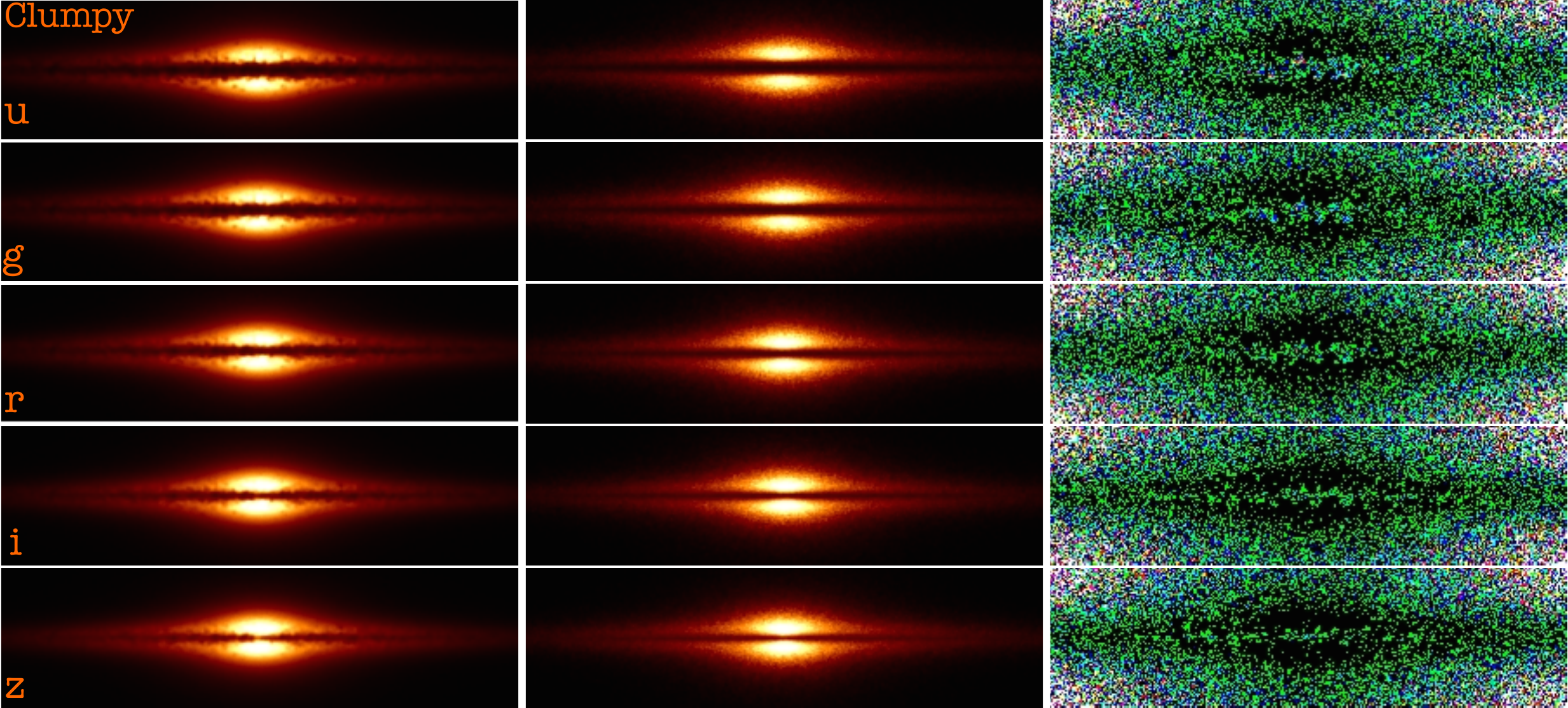}
\includegraphics[width=\textwidth]{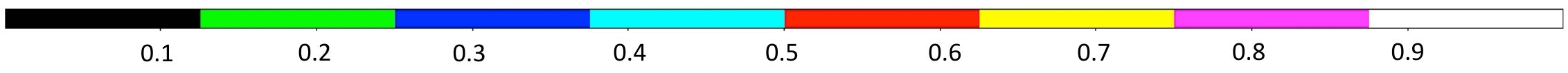}
\caption{Results of the FitSKIRT radiative transfer fits for two basic input models; one with analytical spiral perturbation (upper half) and one with random clumps (lower half). The left column shows the reference images produced by SKIRT in each of the {\em{u}}, {\em{g}}, {\em{r}}, {\em{i}}, and {\em{z}} bands. The middle column shows the corresponding fit obtained with FitSKIRT. The right column contains residual images showing the relative deviation between the fit and the reference image. The colour bar at the bottom presents the scale of the deviation in the residual images.}
\label{FitBasicModels.fig}
\end{figure*}

\begin{table}
\centering
\caption{Model parameter values recovered by the FitSKIRT radiative transfer fits for two basic input models. The third column provides the ``true'' values of the input models, and the fourth and fifth columns list the values recovered for the model with a spiral arm perturbation and a clumpy structure respectively. For a definition of each parameter, see \citet{2014MNRAS.441..869D}.}
\label{FitBasicModels.tab}
\begin{tabular}{cccr@{$\,\pm\,$}lr@{$\,\pm\,$}l}
\hline\hline\\[-0.5ex]
Parameter & Units & Input &
\multicolumn{2}{c}{Spiral} &
\multicolumn{2}{c}{Clumpy} \\[2ex]
\hline \\[-0.5ex]
$h_{R,*}$ & kpc & 4.23 & 4.26 & 0.03 & 4.48 & 0.06\\
$h_{z,*}$ & kpc & 0.51& 0.59 & 0.03 & 0.58 & 0.01 \\
$R_{\text{eff}}$ & kpc & 2.31 & 2.44 & 0.15 & 2.52 & 0.09 \\
$n$ & --- & 2.61& 2.27 & 0.12 & 2.0 & 0.1 \\
$q$ & --- &  0.56 & 0.58 & 0.01 & 0.55 & 0.01 \\
$h_{R,{\text{d}}}$ & kpc & 6.03 & 4.89 & 0.56 & 4.94 & 0.22\\
$h_{z,{\text{d}}}$ & kpc & 0.23 & 0.201 & 0.004 & 0.203& 0.003 \\
$M_{\text{d}}$ & $10^7~M_{\odot}$ & 3.02 & 2.34 & 0.12 & 2.5 & 0.1 \\
$i$ & deg & 90 & 90.01 &  0.01 & 89.96 & 0.02 \\[2ex]
\hline\hline
\end{tabular}
\end{table} 

Finally, the third step in the modelling consists of fitting radiative transfer models to the mock images, as is done for real galaxies. We use the FitSKIRT code for this goal. FitSKIRT \citep{2013A&A...550A..74D, 2014MNRAS.441..869D} is a tool designed to recover the 3D spatial distribution of stars and dust by fitting radiative transfer models to optical images. The code reads any number of images in different bands and simultaneously fits a radiative transfer model to all images. The model can include an arbitrary combination of stellar and dust components. It combines SKIRT with the power of the genetic algorithms-based optimisation library GAlib \citep{Wall96} to perform the actual fitting.  

In this work, we use the same radiative transfer fitting model as the one employed by \citet{2014MNRAS.441..869D} to fit real edge-on spiral galaxies. We fit a smooth axisymmetric model, similar to the underlying model, to the mock images of the perturbed model. The radiative transfer modelling essentially comes down to a strongly nonlinear $\chi^2$-minimalisation in a 21-dimensional parameter space (we consider five parameters describing the stellar geometry, two parameters for the dust geometry, the luminosity of the bulge and disc in each of the five bands, the total dust mass and three projection parameters). For more details, we refer to Section~3.1 of \citet{2014MNRAS.441..869D}. 

The fitting results are presented in Fig.~\ref{FitBasicModels.fig} and Table~\ref{FitBasicModels.tab}. 
The leftmost column in Fig.~\ref{FitBasicModels.fig} represents the original mock images in the five bands, the central column shows the images corresponding to the best fitting model, and the rightmost column presents the residual images, expressed as the relative difference between the surface brightness of the real image and the model fit. The residual frames for the first input model show some discrepancies to the left of the centre of the galaxy, corresponding to a maximum in the spiral arm perturbation. Even in this feature, the relative difference between model and fit is only of the order of 20\%, so this is an excellent fit. For the clumpy input model, the largest discrepancies occur in the {\em{u}} and {\em{g}} bands, where the effects of the dust are most prominent, but in general the quality of the fit is very satisfactory as well.

Table~\ref{FitBasicModels.tab} compares the parameters recovered by the fitting procedure to the corresponding parameters of the input models. For both models, all input parameters are recovered to within 25\%, and often even better. Specifically, the total dust mass is underestimated by less than 25\% in both cases. These results are in line with those reported by \citetalias{2000A&A...353..117M} and \citetalias{2002A&A...384..866M}. 

\section{Input models from hydrodynamical simulations}
\label{HydroInput.sec}

Now that we have shown that we can reproduce the previous results by \citetalias{2000A&A...353..117M} and \citetalias{2002A&A...384..866M}, we can move to the next level, and consider more realistic models for spiral galaxies. The modelling follows the same strategy as we used for the basic input models considered in the previous section. 

\subsection{The input models}

\begin{figure*}
\centering
\includegraphics[width=0.49\textwidth]{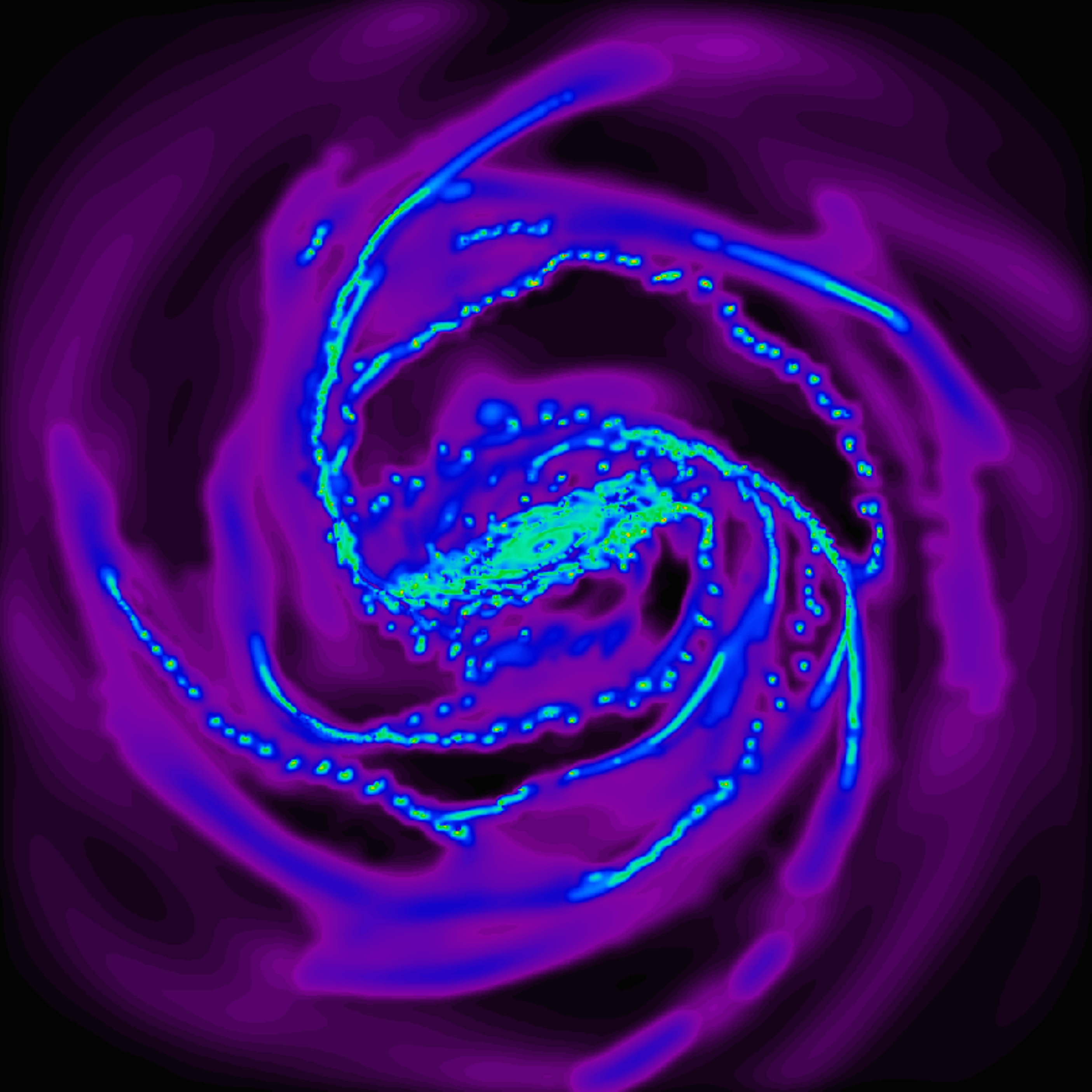}
\includegraphics[width=0.49\textwidth]{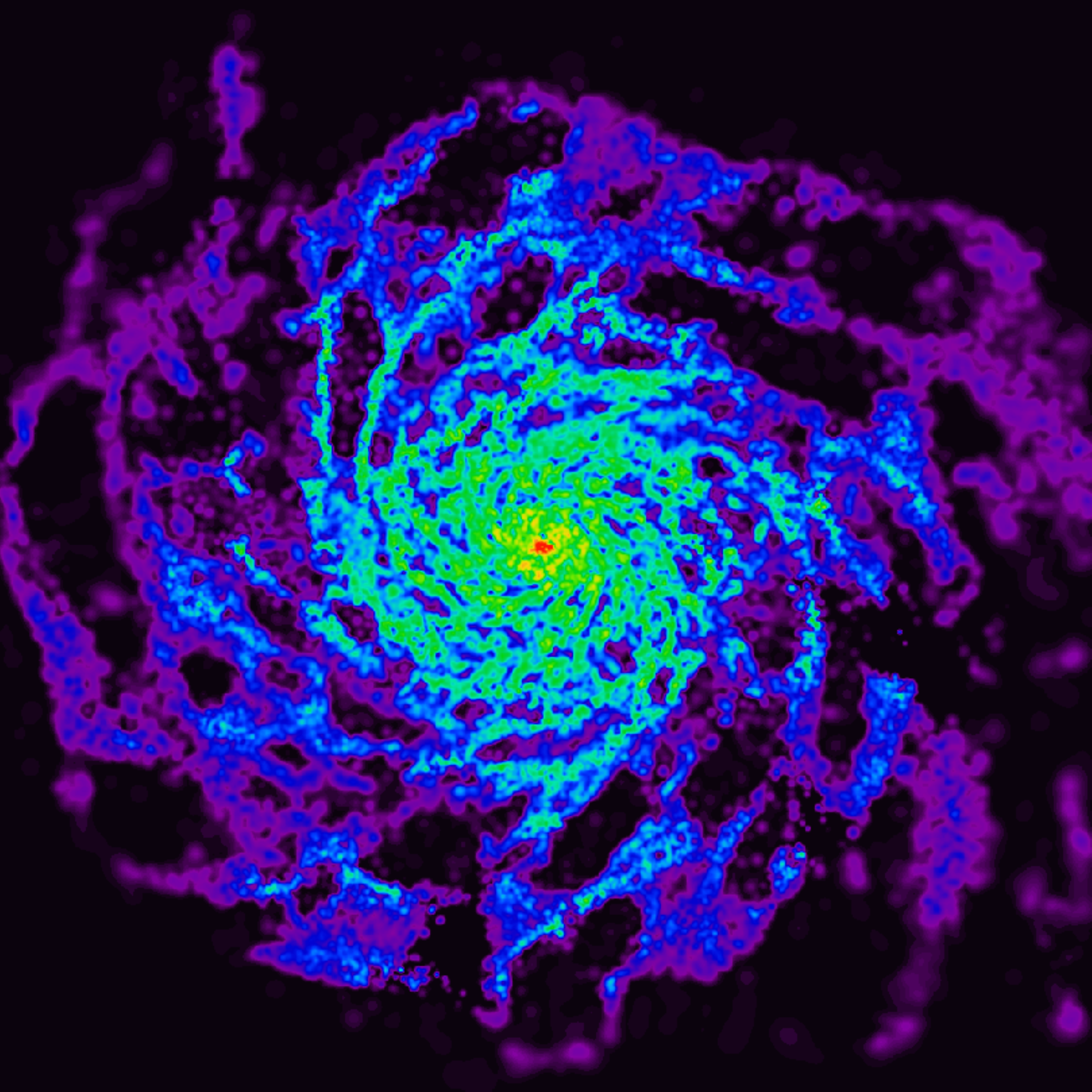}
\includegraphics[width=0.49\textwidth]{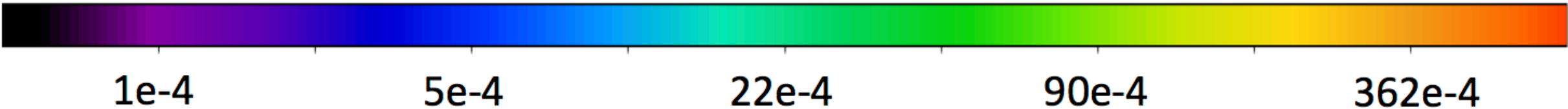}
\includegraphics[width=0.49\textwidth]{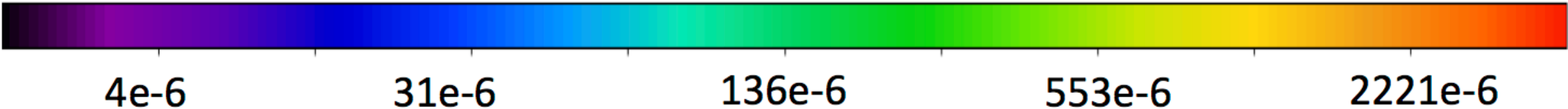}
\caption{A cut through the dust density distribution along the equatorial plane of the R13 galaxy (left panel) and of the Eris galaxy (right panel). The dust density is given in $M_\odot$~pc$^{-3}$.}
\label{DustDensity.fig}
\end{figure*}

In recent years, hydrodynamical simulations have started to successfully reproduce late-type spiral galaxies \citep{2007ApJ...670..237B, 2009ApJ...707L...1B, 2009MNRAS.398..312G, 2011MNRAS.410.1391A, 2011ApJ...742...76G, 2011ApJ...735....1W, 2013MNRAS.428..129S, 2013MNRAS.436.1836R, 2014MNRAS.437.1750M, 2014MNRAS.441..243I}. The spatial resolution of these simulations is sufficient to resolve both large- and small-scale inhomogeneities. We consider snapshots from two different simulations for the analysis in this paper. 

%\subsection{The R13 simulation}

The first input model is taken from the simulation by \citet{2013MNRAS.436.1836R}, hereafter called the R13 simulation. This is a self-consistent hydrodynamical simulation of a Milky Way-like galaxy performed with the Adaptive Mesh Refinement (AMR) code RAMSES \citep{2002A&A...385..337T}. The goal of this simulation was to reproduce an isolated grand-design spiral galaxy and to focus on the structure of the interstellar medium at the highest resolution possible. The simulation resolves the ISM down to scales of 0.05 pc, and includes stellar feedback through photo-ionization, radiative pressure and supernovae. The simulation started with 30 million particles in the dark matter halo and another 30 million particles representing the stars distributed over a bulge, spheroid, thin disc and thick disc. The gaseous disc is represented with 240 million AMR cells varying in size depending on the structure of the ISM. The mass resolution is $160~\mathrm{M}_\odot$ for young star particles and $30~\mathrm{M}_\odot$ for the gas in the most refined cells. The initial components in the simulation are axisymmetric, but non-axisymmetric structures such as a prominent bar and spiral structure are developed quickly because of the instabilities in the velocity distribution. 

We use the snapshot corresponding to the final state of the galaxy, after a run time of 800~Myr. We consider a box with dimensions $20 \times 20 \times 4~{\text{kpc}}^3$, which contains a total stellar mass of $4.4\times10^{10}~M_{\odot}$ and a total gas mass of $3.0\times10^9~M_{\odot}$. The galaxy hosts a prominent bulge and spiral arms, and is characterised by a high star formation rate (SFR); the total SFR for stars younger than 57~Myr is about 7.3~$M_{\odot}$~yr$^{-1}$. The bar has a central area of 1~kpc$^2$ that contains almost no ongoing star formation \citep[see][]{2015MNRAS.446.2468E}, whereas dense star-forming clouds are abundant at the outer regions of the bar. Along the spiral arms, dense clumps of gas and star forming regions have formed with a relatively uniform separation (the so-called ``beads on a string''). This high level of concentration is related to the relatively short simulation run time; continuing the simulation would cause stellar feedback to more evenly spread matter along the disc.

%\subsection{The Eris simulation}

The second input model in our study is taken from the Eris simulation \citep{2011ApJ...742...76G}, one of the most advanced and realistic simulations of the formation of a Milky Way class galaxy. Eris was set up as a zoom-in cosmological simulation, powered by the N-body/SPH GASOLINE code \citep{2004NewA....9..137W}. The simulation follows the formation of a galaxy halo with mass $M_{\text{vir}} = 7.9 \times10^{11}~M_{\odot}$ from $z = 90$ to the present epoch in a full cosmological setting. The target halo is sampled with 26 millions particles divided equally between the dark matter particles and gas particles. Apart from the obvious gravity and hydrodynamical forces, the simulation includes Compton cooling, atomic cooling, metallicity-dependent radiative cooling at low temperatures, the ionising effect of a uniform UV background, star formation and supernova feedback. 

We use the final snapshot corresponding to the present epoch. At $z=0$, Eris is a Milky Way-like galaxy characterised by a beautiful spiral structure and a small bulge in the centre. The structural properties, the mass budget in the various components, and the scaling relations between mass and luminosity are consistent with a host of observational constraints. For our analysis, we consider all particles in a box of $28 \times 28 \times 6~{\text{kpc}}^3$. This box has 7.8 million star particles and 0.25 million gas particles, with a total stellar mass of $3.5 \times 10^{10}~M_{\odot}$ and a total gas mass of $4.3 \times 10^9~M_{\odot}$ (about $5 \times 10^{3}~M_{\odot}$ per star particle and $2 \times 10^4~M_{\odot}$ per gas particle). The smoothing length for the gas particles ranges between 56 and 2455 pc.  

\subsection{Creation of mock images}
\label{mock.sec}

\begin{figure*}
\centering
\includegraphics[width=0.49\textwidth]{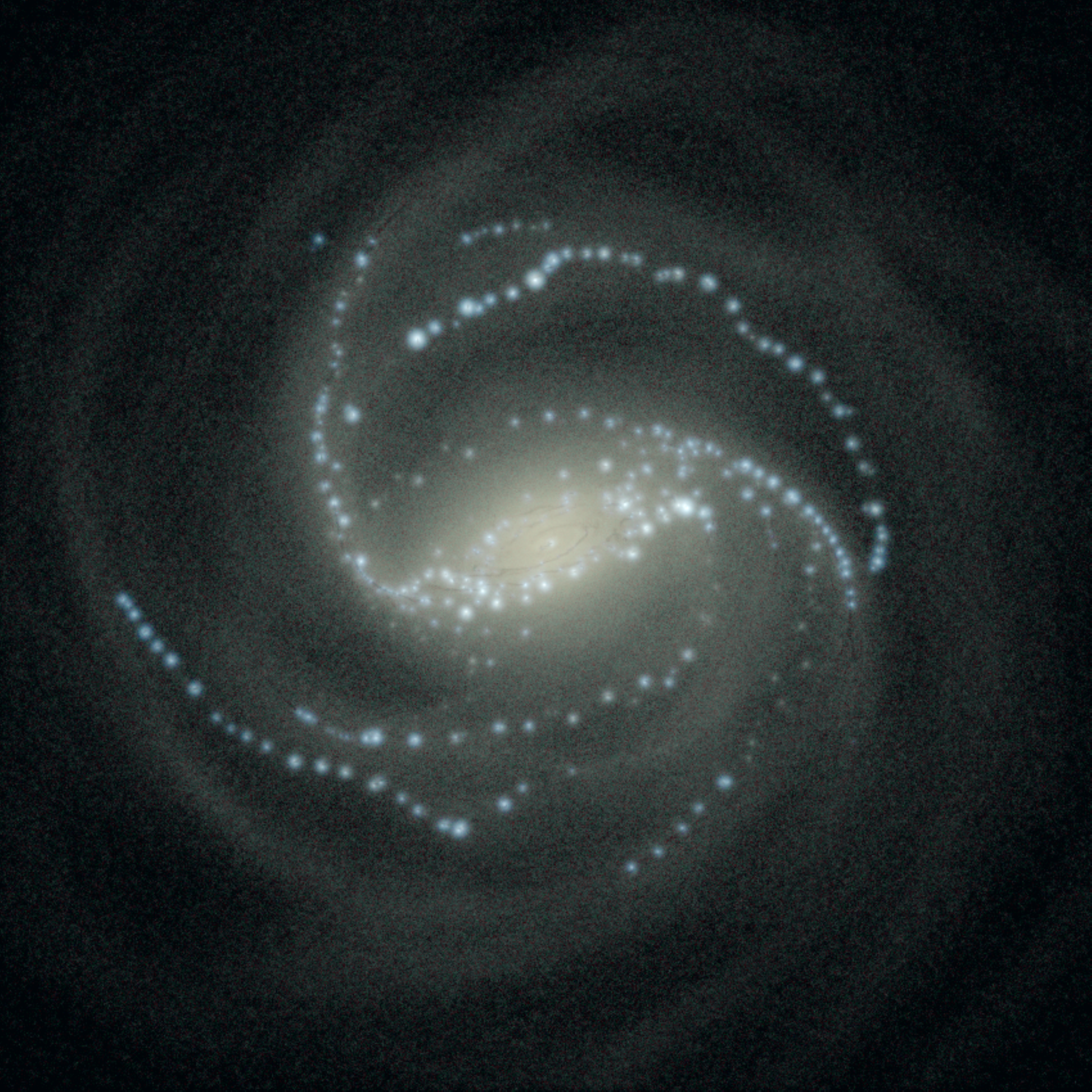}
\includegraphics[width=0.49\textwidth]{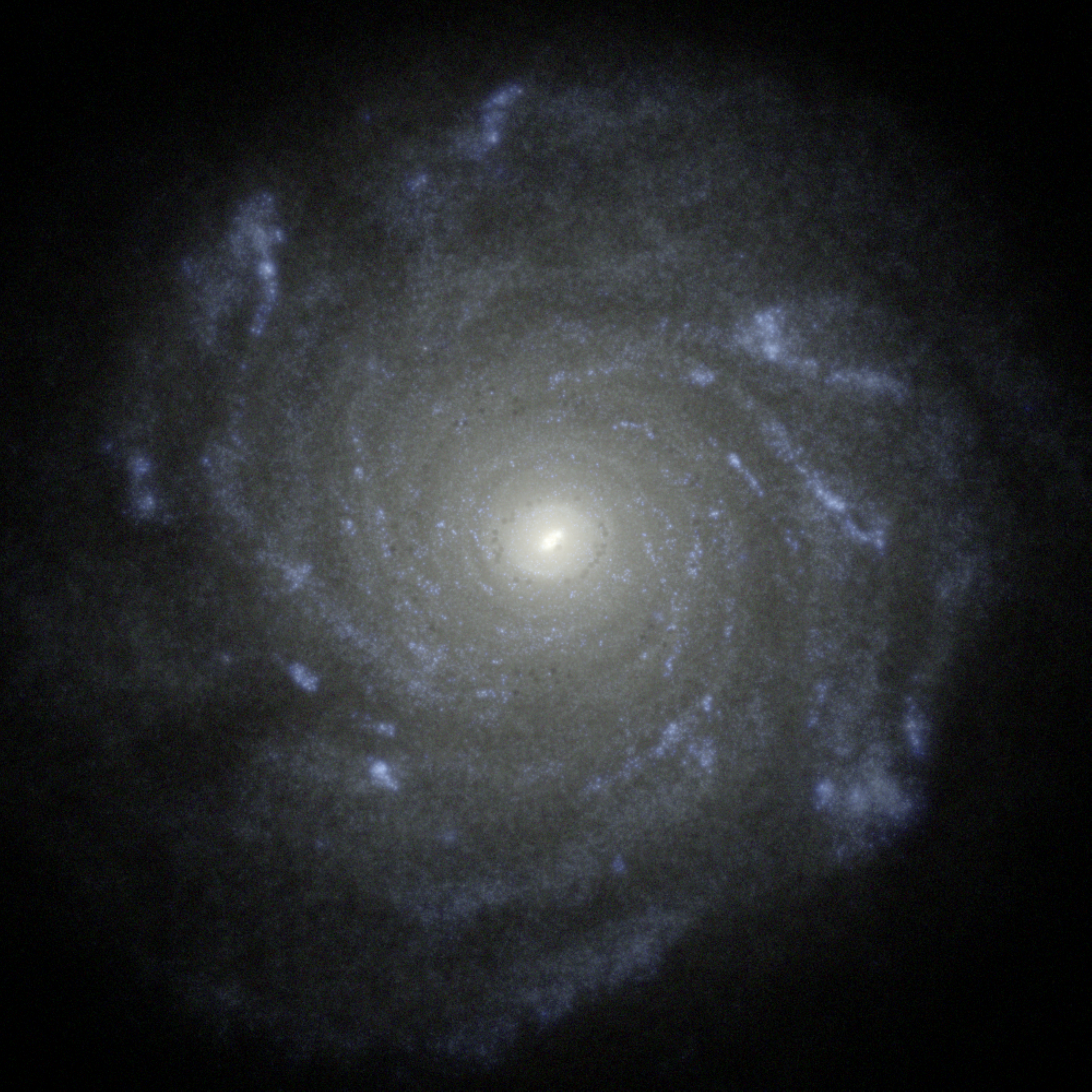}
\caption{Mock face-on views for the R13 galaxy (left panel) and for the Eris galaxy (right panel). The three-colour images are based on the {\em{r}}, {\em{g}} and {\em{u}} band images produced by the SKIRT radiative transfer code.}
\label{FaceOnViews.fig}
\end{figure*}

\begin{figure*}
\centering
\includegraphics[width=0.49\textwidth]{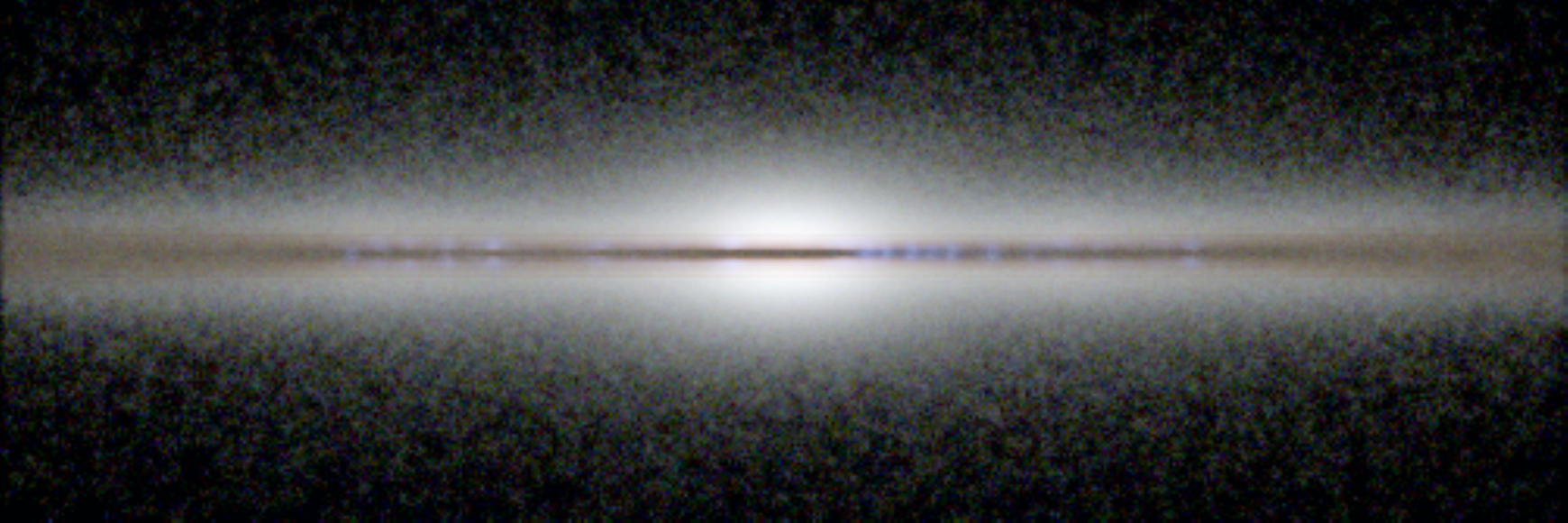}
\includegraphics[width=0.49\textwidth]{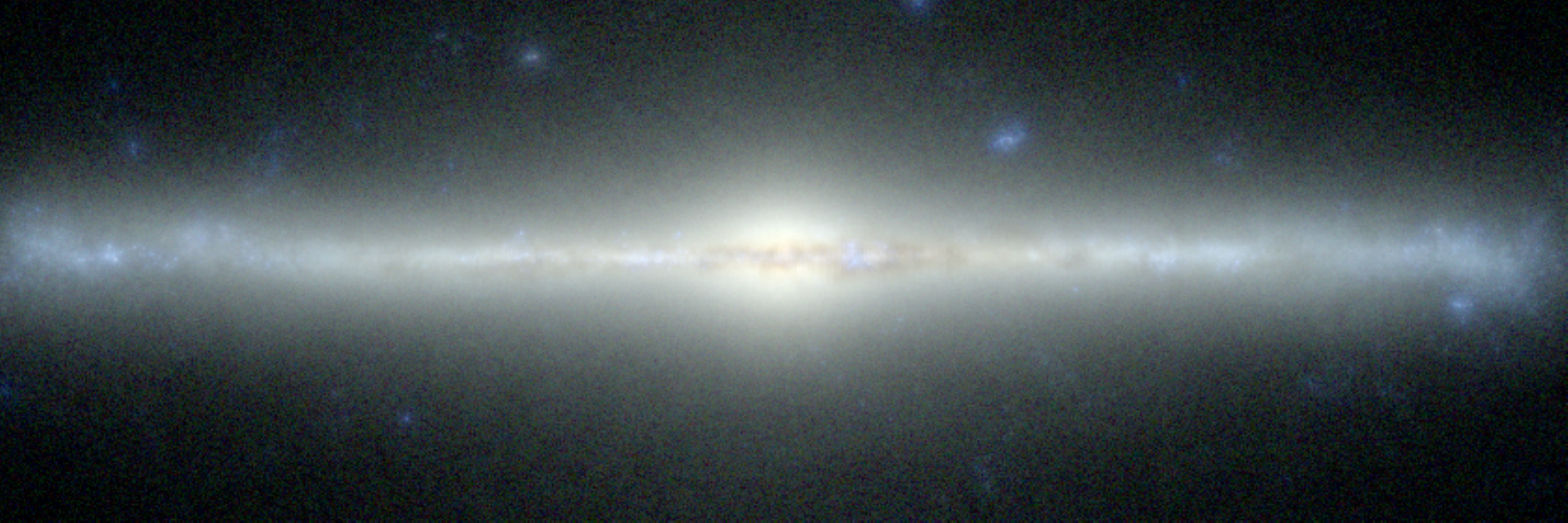}
\caption{Mock edge-on views for the R13 galaxy, looking at the head of the bar (left panel), and for the Eris galaxy (right panel). The three-colour images are based on the {\em{r}}, {\em{g}} and {\em{u}} band images produced by the SKIRT radiative transfer code.}
\label{EdgeOnViews.fig}
\end{figure*}

The second step in the analysis is again the creation of mock images. In order to do so, SKIRT was extended with modules that read the output of hydrodynamical simulations as input for the radiative transfer calculations \citep[see also][]{2015MNRAS.446..521S}. We used the GALAXEV library of simple stellar populations \citep{2003MNRAS.344.1000B} to determine the intrinsic emission of the stars in the simulation. For the optical properties of the dust we use the BARE-GR-S dust model from \citet{2004ApJS..152..211Z}, which is fine-tuned to be consistent with the extinction, diffuse emission and depletion in the Milky Way. 

One particular aspect that needs special care is the determination of the dust density. As the hydrodynamical simulations do not track the dust directly, a recipe needs to be chosen to set the dust density from the properties of the gaseous medium. We use the assumption that a constant fraction of the metals in the ISM is locked up into dust grains. While observations show metallicity gradients across galaxies \citep{2008A&A...484..205D,2012A&A...543A.103P} and local variations in the fraction of metals locked into dust grains \citep{2012MNRAS.422.1263H}, this simplification does not affect our analysis, because we only need to create mock images reflecting a certain dust mass. In particular, we set the 3D dust density using
\begin{equation}
\rho_{\text{dust}} = f_{\text{dust}} \, Z \, \rho_{\text{gas}}
\end{equation}
where $Z$ is the metallicity of the gas, and $f_{\text{dust}}$ is the fraction of metals locked up in dust grains. For the R13 simulation we use $f_{\text{dust}} = 0.3$, the same value as used in the EAGLE simulation \citep{2015MNRAS.446..521S}. For the Eris simulation, we use a slightly larger value, $f_{\text{dust}} = 0.5$, since this galaxy has a relatively low metal content. Both values are within the range suggested by observations, roughly between 0.2 and 0.7 \citep{1998ApJ...501..643D, 2002MNRAS.335..753J, 2012MNRAS.423...38M, 2013A&A...560A..88D, 2013A&A...560A..26Z}. For both simulations, the gas density can be determined from the simulation snapshot. In the R13 simulation, the metallicity is not stored or evolved, and we adopt a fixed solar metallicity. Combined with the fixed value for $f_{\text{dust}}$ this comes down to a constant gas-to-dust ratio of 166, which is in good agreement with the gas-to-dust ratio in the Milky Way and other spiral galaxies \citep{2004ApJS..152..211Z, 2007ApJ...663..866D, 2014A&A...563A..31R}. For the Eris simulation, the gas metallicity is stored for every SPH gas particle, such that the dust density can be calculated. Using a higher value of $f_{\text{dust}}$ boosts the total dust mass, even with the rather low metallicities in this galaxy (ranging from zero up to $Z=0.266$).

Another crucial ingredient for the present simulations is the setup of the grid on which the dust density is defined. SKIRT can handle any 3D geometry, thanks to the efficient use of hierarchical and unstructured grids for the dust medium \citep{2013A&A...554A..10S, 2014A&A...561A..77S, 2013A&A...560A..35C}. For the simulations in this paper, we used a hierarchical $k$-d tree with a resolution up to 0.6 pc for the R13 and 6.8 pc for the Eris simulation. For both simulations, the final grid contains about 1.5 million dust cells. Figure~\ref{DustDensity.fig} shows the dust density in the equatorial plane for both models.

For each of the two input models, we created mock images in the SDSS {\em{ugriz}} bands, in face-on and edge-on configurations (although we only need the edge-on views for our study). Three-colour images based on the {\em{r}}, {\em{g}} and {\em{u}} bands are shown in Figs.~\ref{FaceOnViews.fig} and \ref{EdgeOnViews.fig}. In spite of the complex intrinsic structure of both the stellar and dust distribution apparent in the face-on views, the edge-on morphology seems rather smooth and regular.

\subsection{Radiative transfer modelling}
\label{model.sec}

\begin{figure*}
\centering
\includegraphics[width=1 \textwidth]{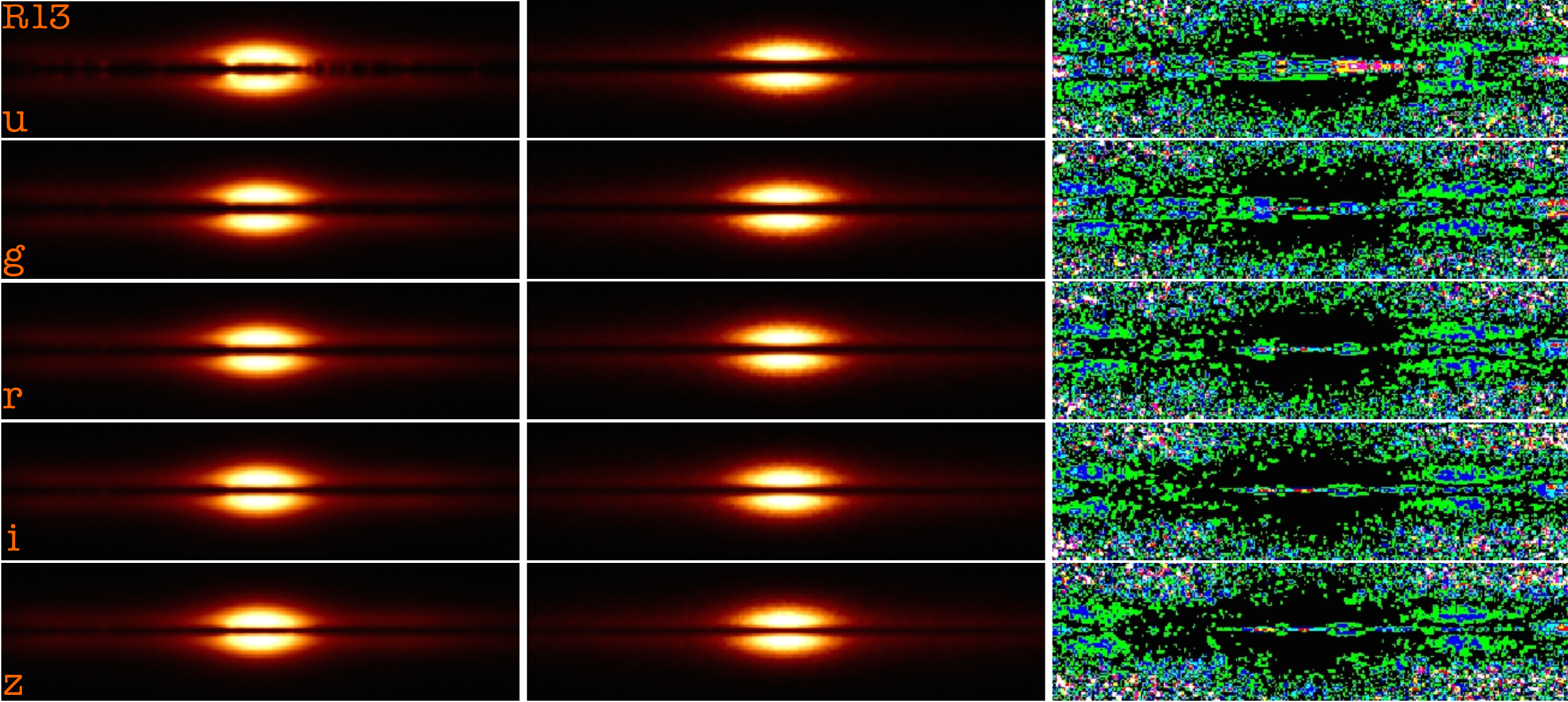}\\[1em]
\includegraphics[width=1 \textwidth]{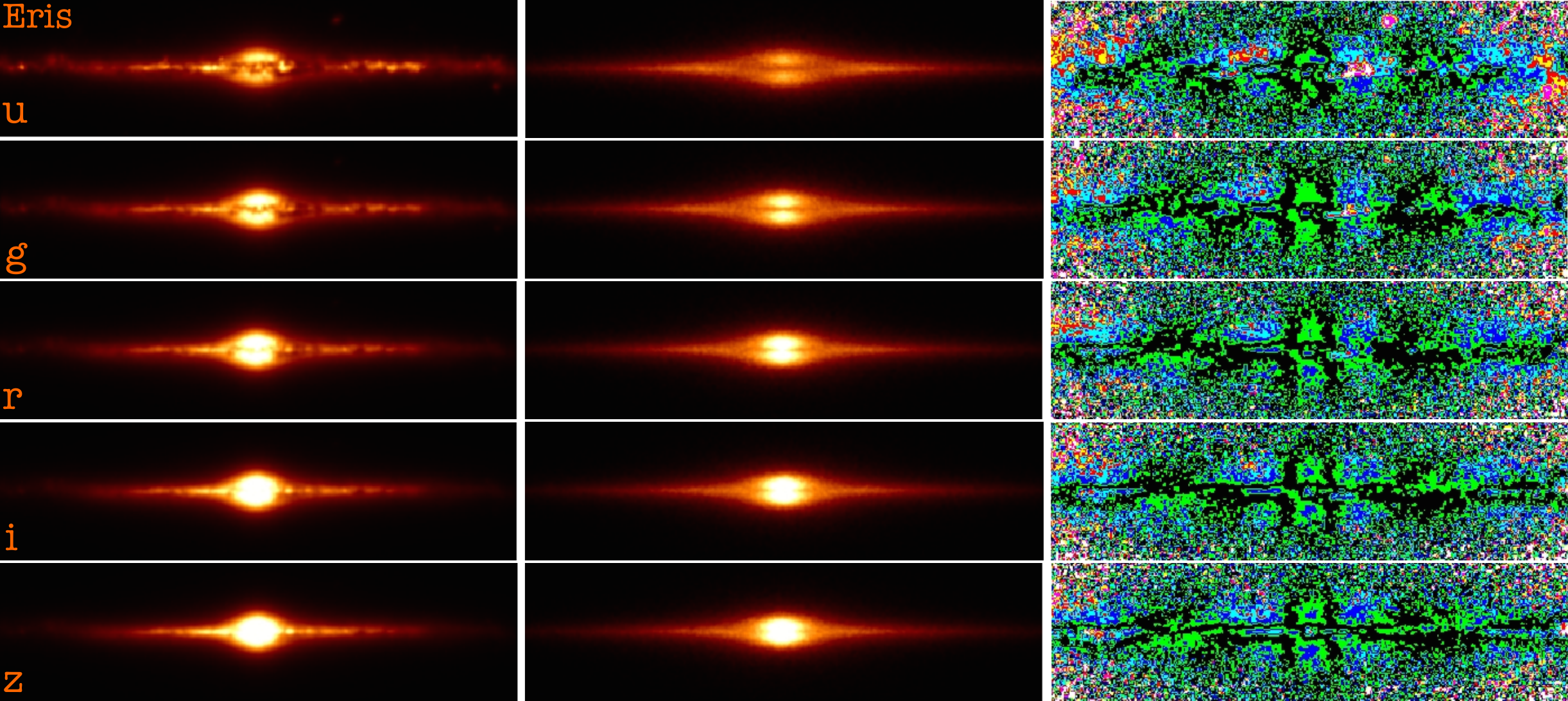}
\includegraphics[width=1 \textwidth]{FitSKIRT_scale.pdf}
\caption{Results of the FitSKIRT radiative transfer fits for the R13 galaxy, looking at the head of the bar (upper half), and for the Eris galaxy (lower half). The left column shows the reference images produced by SKIRT in each of the {\em{u}}, {\em{g}}, {\em{r}}, {\em{i}}, and {\em{z}} bands. The middle column shows the corresponding fit obtained with FitSKIRT. The right column contains residual images showing the relative deviation between the fit and the reference image. The colour bar at the bottom presents the scale of the deviation in the residual images.}
\label{FitRealisticGalaxies.fig}
\end{figure*}

Finally, we fit radiative transfer models to the mock images using FitSKIRT, where we use exactly the same approach as for the basic input models from the previous section. The results for both snapshots are shown in Fig.~\ref{FitRealisticGalaxies.fig} and Table~\ref{FitRealisticGalaxies.tab}. 

The global morphology of each input model is accurately reproduced in the fit, although some features are not fitted completely. In particular, the residual frames for the R13 model show some discrepancies in the dust lane area. This structure is most strongly present in the shortest wavelength bands, but it persists up to the {\em{z}} band. These discrepancies correspond to clumpy structures, both in the stellar distribution (the beads on a string) and in the dust distribution. However, most of the pixels in the residual frames have a discrepancy of less than 30\%. In fact, the central areas of the residual frames for the R13 model shows an obvious similarity to those corresponding to the clumpy disc model (Fig.~\ref{FitBasicModels.fig}, lower panels). For the Eris simulation, the dust lane is not very prominent and it has a discontinuous shape with a lot of clumpy and irregular structures up to the edges of the galaxy, which makes it a hard galaxy to fit. The residual frames show discrepancies around the bulge corresponding to the molecular clouds and star forming regions. 

Table~\ref{FitRealisticGalaxies.tab} lists the most important model parameter values recovered by the radiative transfer fits for each galaxy. To gauge the effect of the large-scale asymmetry in the R13 galaxy due to its prominent bar, we performed two independent fitting procedures based on perpendicular edge-on viewpoints, respectively looking at the head and the side of the bar. Given the dust masses of the input models ($M_\mathrm{d,R13}=1.81\times10^7~M_\odot$; $M_\mathrm{d,Eris}=5.94\times10^6~M_\odot$) and the recovered values in Table~\ref{FitRealisticGalaxies.tab}, it follows that the radiative transfer model underestimates the ``true'' dust mass by a factor of $2.9\pm0.2$ respectively $2.4\pm0.2$ for the R13 galaxy, and by a factor of $4.0\pm0.4$ for the Eris galaxy.

\begin{table}
\centering
\caption{Model parameter values recovered by the FitSKIRT radiative transfer fits for the R13 and Eris galaxies. For the R13 galaxy, we list two sets of parameters (head and side). They correspond to two independent radiative transfer fits, based on perpendicular edge-on viewpoints, respectively looking at the head and the side of the bar. For a definition of each parameter, see \citet{2014MNRAS.441..869D}.}
\label{FitRealisticGalaxies.tab}
\begin{tabular}{ccr@{$\,\pm\,$}lr@{$\,\pm\,$}lr@{$\,\pm\,$}l}
\hline\hline\\[-0.5ex]
Par. & Units &
\multicolumn{2}{c}{R13 (head)} &
\multicolumn{2}{c}{R13 (side)} &
\multicolumn{2}{c}{Eris} \\[2ex]
\hline \\[-0.5ex]
$h_{R,*}$ & kpc & 4.17 & 0.38 & 3.57 & 0.39 & 4.14 & 0.13\\
$h_{z,*}$ & kpc & 0.43 & 0.01 & 0.40 & 0.01 & 0.37 & 0.01\\
$R_{\text{eff}}$ & kpc & 0.93 & 0.04 & 1.12 & 0.08 & 0.49 & 0.07\\
$n$ & --- & 0.64 & 0.03 & 0.80 & 0.04 & 6.71 & 0.48 \\
$q$ & --- & 0.47 & 0.01 & 0.40 & 0.03 & 0.56 & 0.03\\
$h_{R,{\text{d}}}$ & kpc &  9.47 & 0.69 & 4.23 & 0.80 & 0.35 & 0.05\\
$h_{z,{\text{d}}}$ & kpc &  0.064 & 0.005 & 0.091 & 0.006 & 0.095 & 0.008\\
$M_{\text{d}}$ & $10^6~M_{\odot}$ & 6.26 & 0.24 & 7.68 & 0.33 & 1.48 & 0.18\\
$i$ & deg & 89.96 &  0.02 &  89.96 & 0.02 & 90.04 & 0.16\\[2ex]
\hline\hline
\end{tabular}
\end{table}

\section{Discussion}
\label{Discussion.sec}

The previous work by \citetalias{2000A&A...353..117M} and \citetalias{2002A&A...384..866M} cited in Sect.~\ref{Introduction.sec}, and our results presented in Sect.~\ref{BasicInput.sec}, indicate that fitting a smooth galaxy model to a certain class of basic, quasi-analytical input models recovers the intrinsic dust mass of the input model to within 40\% or better. In fact, as evidenced by the close correspondence between recovered and input values listed in Table~\ref{FitBasicModels.tab}, a logarithmic spiral arm perturbation or a clumpy dust distribution seem to have only a modest effect on the structural parameters observed in an optical edge-on view of a disc galaxy, including the derived dust mass. 

In contrast, our results presented in Sect.~\ref{HydroInput.sec} indicate that fitting a smooth galaxy model to more realistic galaxy models, obtained from high-resolution hydrodynamical simulations, underestimates the intrinsic dust mass of the input model by a factor of about three. This is a tantalising result, especially since a factor of roughly the same magnitude has been found in energy balance studies of real edge-on spiral galaxies \citep{2000A&A...362..138P,2001A&A...372..775M, 2004A&A...425..109A, 2005A&A...437..447D, 2010A&A...518L..39B, 2012MNRAS.427.2797D, 2012MNRAS.419..895D}.

It is tempting to conclude that the level of dust underestimation is driven by the fundamental differences between the input models. The models in \citetalias{2000A&A...353..117M} and \citetalias{2002A&A...384..866M}, and in our Sect.~\ref{BasicInput.sec}, are derived from well-behaved, smooth disc models by applying a relatively modest perturbation. For example, the spiral arm perturbation is fully analytical and cancels out exactly when averaged over azimuth. The galaxy models constructed from hydrodynamical simulation snapshots, presented in Sect.~\ref{HydroInput.sec}, feature much more realistic inhomogeneities at a wide range of scales, from large-scale bars and spiral arms to parsec-sized clumps and filaments. These structural complexities may very well be responsible for a higher level of dust underestimation in the radiative transfer fits.

In other words, our modelling suggests that the complex and inhomogeneous structure of galaxies can hide up to three times more dust than is ``observed'' when the optical images are fitted with smooth axisymmetric models. FIR/submm observations of several spiral galaxies also imply a factor of three times more dust than visible in the optical; this correspondence suggest that the inhomogeneous structure of the ISM possibly is the source of the dust energy balance problem. The recent work by \citet{2014A&A...571A..69D} supports this hypothesis. They performed a detailed panchromatic radiative transfer modelling of the face-on galaxy M51 with a model that includes the complex geometry as derived from the FUV attenuation map. The model self-consistently reproduces the surface brightness images from UV to submm wavelengths. The face-on analysis is of course less affected by optical depth effects along the line of sight, which may have contributed to this result as well.

We must be careful not to jump to conclusions. First, while a typical factor of about three is found by several teams for different edge-on spiral galaxies \citep{2000A&A...362..138P,2001A&A...372..775M, 2004A&A...425..109A, 2005A&A...437..447D, 2010A&A...518L..39B, 2012MNRAS.427.2797D, 2012MNRAS.419..895D}, this is by no means an ubiquitous feature. This was shown most recently by \citet{DeGeyter2015}, who performed the same fitting procedure as presented in this paper on two edge-on spiral galaxies from the sample analysed in \citet{2014MNRAS.441..869D}. For one of the two galaxies, a typical factor-of-three discrepancy is observed between the best fitting FitSKIRT model and the observed FIR/submm SED, whereas for the other galaxy the FitSKIRT model accurately describes the observed spectrum both in absolute values and shape.

Secondly, even for those galaxies in which a dust energy balance problem is encountered, it is up to debate whether this can be ascribed to the same physical scenario. For example, the dust emission excess in the Sombrero galaxy is shown to be compatible with an additional unresolved cold dust reservoir \citep{2012MNRAS.419..895D}, whereas the excess emission in the edge-on spiral UGC\,4754 is rather compatible with an additional warmer component, such as expected when linked to recent star formation \citep{2010A&A...518L..39B, 2011ApJS..196...22B}.

Finally, our present study is based on just two simulated spiral galaxies, and each of them has its strengths and weaknesses. In both galaxies we recognise the structures and morphologies of real galaxies, including spiral arms, bars, bulges, star forming regions, and compact clumps. However, the R13 galaxy contains star forming regions that look somewhat artificial, and the galaxy's dust lane is very thin and extended; and the Eris galaxy has a faint and rather fuzzy dust lane that is not visible in the {\em{r}}, {\em{i}} and {\em{z}} bands, which is atypical for real galaxies.

\section{Conclusion}
\label{Conclusion.sec}

We set out to shed light on the dust energy balance problem in edge-on spiral galaxies by performing the radiative transfer fitting procedure that has been used previously for studying real galaxies on snapshots obtained from hydrodynamical simulations. These simulated galaxies feature a more realistic inhomogeneous structure than typical quasi-analytical models, and their ``true'' dust mass is known.

We used two simulated Milky Way-like galaxies as input models. The R13 simulation is a self-consistent hydrodynamical simulation performed with the AMR code RAMSES. The Eris simulation is a zoom-in cosmological simulation performed by the N-body/SPH GASOLINE code. We used our radiative transfer code SKIRT to create mock observational images in the SDSS {\em{ugriz}} bands for an edge-on view of both galaxies, and we fitted the parameters of a basic, smooth disc galaxy model to these images with our radiative transfer fitting code FitSKIRT. We found that, for both galaxies, the dust mass is underestimated by a factor of about three. This result is strikingly close to what has been found in previous work by several teams who performed similar analyses for real edge-on spiral galaxies.

In contrast, previous studies have shown that fitting a smooth disc galaxy model to a modestly perturbed, quasi-analytical model (including structures such as spiral arms or dust clumps) can properly recover the ``true'' dust mass of the input model. To eliminate the possibility that our fitting procedure would behave differently, we repeated our analysis for such basic input models, using exactly the same procedure as for the more realistic input models. We found that our fits could indeed recover the proper dust mass within a narrow margin.

These results suggest that the level of dust underestimation is driven by the fundamental differences between the input models, implying that the complex and inhomogeneous structure of galaxies can hide up to three times more dust than is ``observed'' when the optical images are fitted with smooth axisymmetric models. Although our analysis is too anecdotal to be conclusive, we would still argue that this effect may help explain the dust energy balance problem, at least in part and for certain types of galaxies.

\begin{acknowledgement}
We thank Aleksandra Sokolowska for providing support on handling the Eris simulation data. WS acknowledges the support of Al-Baath University and The Ministry of High Education in Syria in the form of a research grant. MB, GDG and IDL gratefully acknowledge the support from the Flemish Fund for Scientific Research (FWO-Vlaanderen). This work fits in the CHARM framework (Contemporary physical challenges in Heliospheric and AstRophysical Models), a phase VII Interuniversity Attraction Pole (IAP) programme organised by BELSPO, the BELgian federal Science Policy Office. Part of the computational resources and services used in this work were provided by the VSC (Flemish Supercomputer Center), funded by the Hercules Foundation and the Flemish Government -- department EWI.
\end{acknowledgement}

\bibliographystyle{aa} % style aa.bst
\bibliography{DustEnergyBalance}

\end{document}